# Inhibition of light tunneling for multichannel excitations in longitudinally modulated waveguide arrays

Valery E. Lobanov,[1,2] Victor A. Vysloukh,[3] and Yaroslav V. Kartashov[1]

[1]*ICFO-Institut de Ciencies Fotoniques, and Universitat Politecnica de Catalunya, Mediterranean Technology Park, 08860 Castelldefels (Barcelona), Spain*

[2]*Faculty of Physics, Lomonosov Moscow State University, Leninskie Gory, 119991 Moscow, Russia*

[3]*Departamento de Fisica y Matematicas, Universidad de las Americas – Puebla, Santa Catarina Martir, 72820, Puebla, Mexico*

We consider evolution of multichannel excitations in longitudinally modulated waveguide arrays where refractive index either oscillates out-of-phase in all neighboring waveguides or when it is modulated in phase in several central waveguides surrounded by out-of-phase oscillating neighbors. Both types of modulations allow resonant inhibition of light tunneling, but only the modulation of latter type conserves the internal structure of multichannel excitations. We show that parameter regions where light tunneling inhibition is possible depend on the symmetry and structure of multichannel excitations. Antisymmetric multichannel excitations are more robust than their symmetric counterparts and experience nonlinearity-induced delocalization at higher amplitudes.

Optical structures with periodic transverse modulation of the refractive index provide unique opportunities for precise control of the propagation dynamics of light, allowing, e.g., engineering of diffraction [1,2]. Biperiodic modulations along both transversal and longitudinal directions open new routes for diffraction management and make possible a variety of new phenomena including formation of diffraction-managed solitons [3,4], dragging of laser beams [5,6], periodic shape transformations or Rabi oscillations [7,8], and parametric amplification of soliton swinging [9], just to mention a few. One of the most promising features of longitudinally modulated waveguide arrays is the possibility of discrete diffraction suppression (or light tunneling inhibition), even in the linear regime. Such suppression is a resonant effect, occurring only for specific set of modulation amplitudes and frequencies. Diffraction suppression was demonstrated in periodically curved arrays [10–15], in arrays with oscillat-ing widths of channels [16], and in lattices with longitudinally oscillating refractive index [14,17–21]. Most of the effects mentioned above were demonstrated for simplest single-site excitations or for broad nodeless beams.

The specific features of tunneling inhibition in linear and weakly nonlinear regimes for higher-order complex modes incorporating multiple bright spots were not studied yet. The stability of higher-order modes was investigated only in strongly nonlinear regime when localization occurs due to solitonic effect (i.e. when input amplitude is already sufficiently high for formation of localized states even without longitudinal modulation of system parameters) in deeply modulated lattices with defocusing nonlinearities [22] and in periodically bending waveguide arrays [23]. A detailed analysis of tunneling inhibition for multichannel excitation would be especially interesting in structures where diffraction is not fully suppressed [14], e.g., in arrays with oscillating widths of channels or in lattices with longitudinally oscillating refractive index.

In this work we show the possibility of tunneling inhibition for higher-order excitations in waveguide arrays with longitudinal refractive index modulations. We put forward a novel type of refractive index modulation that allows inhibition of tunneling and simultaneously preserves internal mode structure. The dependence of parameter regions where light tunneling inhibition is possible on the symmetry and structure of multichannel excitations is also discussed.

For analysis we use the full continuous model since simplified discrete model allows only a qualitative description of the phenomenon. In particular, in the frames of latter model it is hardly possible to take into account mode width oscillations arising due to longitudinal refractive index modulation as well as radiative losses. When the refractive index modulation depth increases such oscillations become considerable and may substantially affect tunneling inhibition dynamics and corresponding resonant frequencies. Our model is based on the nonlinear Schrödinger equation for the dimensionless field amplitude $q$, governing the propagation of light beam along the $\xi$ axis of waveguide array with longitudinally modulated refractive index:

$$i\frac{\partial q}{\partial \xi} = -\frac{1}{2}\frac{\partial^2 q}{\partial \eta^2} - pR(\eta,\xi)q - |q|^2 q. \qquad (1)$$

Here, $\eta$ and $\xi$ are the normalized transverse and longitudinal coordinates, while $p$ stands for the refractive index contrast of the individual waveguide. The refractive index profile of the lattice is given by $R(\eta,\xi) = \sum_{m=-\infty}^{+\infty}[1 + F(m)\mu\sin(\Omega\xi)]G[\eta - (m+1/2)w_{\mathrm{s}}]$, where $w_{\mathrm{s}}$ is the separation between waveguides, $\mu$ is the longitudinal modulation amplitude, $\Omega$ is the modulation frequency, the function $G(\eta) = \exp(-\eta^6/w_\eta^6)$ describes profile of individual waveguides with widths $w_\eta$, while the function $F(m)$ determines the type of refractive index modulation. Qualitatively similar results can be obtained also for other waveguide shapes.

Further we consider two types of longitudinal refractive index modulation (see Fig. 1). In the first case (termed "out-of-phase" modulation) the refractive index oscillates out-of-phase in all waveguides of the array that corresponds to $F(m) = (-1)^m$ [see Fig. 1(b)]. Such modulation, as proposed in [16], was used before to demonstrate inhibition of light tunneling for single-channel excitations experimentally [19]. In the second case (further called "in-phase" modulation) the refractive index oscillates in-phase in the selected group of several excited waveguides, while in all other waveguides surrounding the selected group it oscillates out-of-phase. For instance, for simplest two-channel excitation in the form of symmetric or antisymmetric two-hump modes, the refractive index oscillates in-phase in two central channels, i.e., $F(m) = (-1)^{m+1}$ for $m < 0$, and $F(m) = (-1)^m$ for $m \geq 0$ [see Fig. 1(c)]. In this work we focus on investigation of tunneling inhibition for multichannel excitations and consider the input conditions of the form $q|_{\xi=0} = Aw(\eta)$, where $A$ is the input amplitude, while $w(\eta)$ represents the profiles of various linear guided modes of a group of several isolated single-mode waveguides [thus, in simplest case of two-channel excitations we use as an input symmetric and antisymmetric modes of two guides depicted in Fig. 2(d)]. We further set $w_\eta = 0.3$, $w_s = 1.8$, and $p = 8.7$. To characterize the efficiency of tunneling inhibition we introduce distance-averaged power fraction trapped in several central excited channels. For two-channel excitations distance-averaged power fraction is given by:

$$U_\mathrm{m} = L^{-1} \int_0^L d\xi \int_{-w_s}^{w_s} |q(\eta,\xi)|^2 \, d\eta \bigg/ \int_{-w_s}^{w_s} |q(\eta,0)|^2 \, d\eta, \qquad (2)$$

where $L$ is the propagation distance. We also monitored the distance-averaged power fraction trapped in each of the excited waveguides.

First, we demonstrate that light tunneling inhibition is possible for multichannel excitations even in linear regime for both "in-phase" and "out-of-phase" types of longitudinal modulation. In order to do so we calculated distance-averaged power fraction $U_\mathrm{m}$ as a function of modulation frequency $\Omega$. It is instructive to normalize the modulation frequency to the beating frequency of unmodulated linear coupler $\Omega_\mathrm{b} = 2\pi/T_\mathrm{b}$, where $T_\mathrm{b}$ is a beating period. For our set of parameters one has $T_b = 52.8$. In all cases considered the propagation distance was $L = 4T_b$.

Figures 2(a) and 2(b) show resonance curves $U_m(\Omega)$ for antisymmetric two-channel excitation [curve "a" in Fig. 2(d)] in "out-of-phase" and "in-phase" modulated waveguide arrays. One can see that inhibition of light tunneling takes place for both types of modulation for the specific resonant frequencies, but resonances are much narrower in the case of "out-of-phase" modulation. This indicates that "in-phase" modulation is more favorable for multichannel excitations, since it allows inhibition of tunneling within broader range of modulation frequencies. The resonance frequencies are practically the same for both types of modulation. However, in the case of "in-phase" modulation the distance-averaged power fractions trapped in each of two waveguides and in both excited waveguides exactly coincide that indicates the absence of energy exchange between two excited central waveguides. In contrast, upon "out-of-phase" modulation distance-averaged power fractions in different channels are different and even acquire maximal values at slightly different modulation frequencies that do not coincide with a common resonance frequency defined for a pair of central waveguides. Notice that change $\mu \to -\mu$ (or $\pi$-shift of phase of modulation) does not affect position of common resonance, but it results in exchange of resonances for individual channels. It should be mentioned that due to definition of $U_m$ the values of distance-averaged power fraction in each of the excited waveguide can exceed unity [see, e.g., curve 2 in Fig. 2(c)], although their half-sum for each particular frequency of longitudinal modulation coincides with the value of $U_m$ defined for the pair of waveguides that is always smaller than 1. This shows that "out-of-phase" modulation allows achieving inhibition of tunneling for multichannel excitations, but it also results in energy exchange between excited waveguides and distortion of the internal structure of input "collective" mode. A similar picture was encountered for excitations of three and larger number of waveguides.

We also studied inhibition of light tunneling for multichannel excitations in nonlinear regime. While for single-site excitations the width of primary resonance is a monotonically growing function of the input amplitude (as it was shown recently [20]), we found that for multichannel excitations the situation might be dramatically different. To illustrate this, we calculated resonance curves for different input amplitudes for both symmetric and antisymmetric modes for different types of longitudinal modulation. Surprisingly, in the case of "in-phase" modulation for antisymmetric excitation the width of resonance first decreases with growth of amplitude, reaches certain minimal value at intermediate amplitudes, and then starts increasing [Fig. 3(a)]. At the same time, the width of resonance for symmetric excita-tion in a system with such type of longitudinal modulation grows monotonically with amplitude [Fig. 3(b)]. The corresponding dependencies of resonance width $\delta\Omega$ defined at the level $0.9U_{\max}$ (here $U_{\max}$ is the distance-averaged power fraction in primary resonance) on amplitude are presented in Fig. 3(c). The dependence $\delta\Omega(A)$ for the symmetric excitation is approximately parabolic for sufficiently high amplitudes. Interestingly, in array with "out-of-phase" longitudinal refractive index modulation the resonance width increases with $A$ monotonically for both symmetric and antisymmetric modes [Fig. 3(d)]. Moreover, while for "in-phase" modulation the resonance widths for symmetric and antisymmetric modes at $A \to 0$ are different, in the case of "out-of-phase" modulation the dependencies $\delta\Omega(A)$ for both modes are very close for any amplitude $A$ [hence, in Fig. 3(d) we show only dependence for symmetric mode]. All features mentioned above imply that the band of modulation frequencies where tunneling inhibition is possible is rather sensitive to the type of longitudinal modulation and to the structure of input collective mode.

As in the case of single-site excitations further growth of the input amplitude results in nonlinearity-induced delocalization of higher-order modes [see Fig. 4(a) that shows dependence of distance-averaged power fraction on input amplitude for both symmetric and antisymmetric modes for "in-phase" modulation]. In the regime of nonlinearity-induced delocalization the light beam may spread considerably across the waveguide array. Interestingly, for antisymmetric modes and "in-phase" modulation the output intensity pattern remains symmetric even in the delocalization regime, while for symmetric excitations one observes remarkable asymmetries in output intensity distributions [compare Figs. 5(c) and 5(e)]. If the amplitude increases even further, the nonlinearity-induced delocalization is gradually replaced by the localization due to the soliton-type mechanism. We calculated the dependence of the critical amplitude $A_{\text{del}}$ at which the distance-averaged power fraction in the excited channels decreases below the level of $0.9U_{\max}$ (here $U_{\max}$ is the distance-averaged power fraction in the primary resonance at $A \to 0$) and the mode experiences nonlinearity-induced delocalization on the depth of longitudinal modulation $\mu$. We found that $A_{\text{del}}$ monotonically increases with $\mu$ [see Fig. 4(b)], while symmetric mode experiences delocalization at smaller values of input amplitudes than its antisymmetric counterpart as it is shown in Fig. 4(a). This feature is rather surprising taking into account almost identical asymptotical behavior of tails of symmetric and antisymmetric modes, which intuitively suggests that in the unmodulated arrays such modes would experience a similar rate of discrete diffraction [see Fig. 5(a) showing diffraction pattern for antisymmetric mode in unmodulated array]. One therefore may conclude that antisymmetric modes are more robust in the longitudinally modulated arrays. However, the fact that symmetric modes experience delocalization at smaller amplitudes is not connected with symmetry-breaking instability analogous to instabilities of even solitons in waveguide arrays. Thus, energy exchange between two central waveguides in the case of symmetric mode (which might serve as an indication of development of such kind of instability) begins at the amplitude value that is even smaller than amplitude of overall nonlinearity-induced delocalization $A_{\text{del}}$. Importantly, this amplitude does not depend on the longitudinal modulation depth, while $A_{\text{del}}$ does depend on $\mu$. Although this instability does not result in overall delocalization it may cause asymmetry in output intensity distributions for initially symmetric input modes.

The representative propagation dynamics for two-channel excitations is presented in Fig. 5. Figure 5(a) demonstrates diffraction of antisymmetric mode in unmodulated waveguide array [Fig. 1(a)]. Figures 5(b) - 5(d) illustrate propagation of antisymmetric modes in "in-phase" modulated waveguide array when modulation frequency corresponds to primary linear resonance. In practically linear regime at $A = 0.01$ one observes localization [Fig. 5(b)] (it should be stressed that although longitudinal refractive index modulation in such structure does not result in 100% inhibition of tunneling the distance-averaged power fraction in resonance is still very close to unity, so that coupling to other channels is not visible at distances considered here); nonlinearity-induced delocalization takes place at intermediate amplitudes [$A = 1.59$, Fig. 5(c)]; the relocalization can be observed at sufficiently high amplitude [$A = 2.00$, Fig. 5(d)]. Figures 5(e) and 5(f) compare propagation of symmetric and antisymmetric modes with equal amplitudes in such setting. While antisymmetric mode is still localized [Fig. 5(f)], its symmetric counterpart already experiences delocalization [Fig. 5(e)] in agreement with results presented in Fig. 4(b).

Finally, Fig. 6 illustrates that longitudinal refractive index modulation allows inhibition of light tunneling not only for two-channel excitations, but also for complex multichannel ones. As one can see from Fig. 6(c), the antisymmetric fourth-order mode perfectly preserves its structure in the "in-phase" modulated lattice [see Fig. 6(a) for corresponding lattice profile], while in the case of "out-of-phase" modulation [see corresponding lattice in Fig. 1(b)] the structure of this mode is strongly distorted due to energy exchange between adjacent waveguides [Fig. 6(b)], although in both cases the coupling between the central group of excited guides and surrounding array is inhibited.

Summarizing, we showed that inhibition of light tunneling can be achieved not only for single-site excitations, but also for complex multichannel excitations in properly modulated waveguide arrays. Thus, specific "in-phase" modulation is found to be more favorable for modes with multiple humps than conventional "out-of-phase" modulation when refractive index in all waveguides of array oscillates out-of-phase. We also showed that the band of modulation frequency where tunneling inhibition is possible and the delocalization amplitude value are sensitive to the type of longitudinal modulation and the structure of input collective mode.

The work of V.E.L. was supported, in part, by the Ministry of Science and Education of Russian Federation (Program Leading Scientific Schools, project NSh-671.2008.2) and the Russian Foundation for Basic Research (projects 08-02-00717 and 09-02-01028).

# Figure captions

Figure 1. Unmodulated (a), "out-of-phase" modulated (b), and "in-phase" modulated (c) waveguide arrays.

Figure 2. Theoretically calculated $U_\mathrm{m}$ value in both central waveguides versus $\Omega/\Omega_\mathrm{b}$ at $\mu=0.2$ and $A=0.01$ for antisymmetric mode in (a) "out-of-phase" modulated and (b) "in-phase" modulated waveguide arrays. (c) $U_\mathrm{m}$ versus $\Omega/\Omega_\mathrm{b}$ in right (curve 1) and left (curve 2) waveguides for antisymmetric mode in "out-of-phase" modulated array at $\mu=0.2$ and $A=0.01$. Dashed lines in (a)-(c) indicate $U_\mathrm{m}=1$ level. (d) Profiles of symmetric and antisymmetric modes of two waveguides.

Figure 3. $U_\mathrm{m}$ versus $\Omega/\Omega_\mathrm{b}$ for (a) antisymmetric mode with amplitude $A=0.01$ (curve 1), $0.35$ (curve 2), $0.42$ (curve 3) and (b) symmetric mode with amplitude $A=0.01$ (curve 1), $0.25$ (curve 2), $0.40$ (curve 3) in "in-phase" modulated waveguide array at $\mu=0.2$. Dashed lines indicate $U_\mathrm{m}=1$ level. (c) Resonance curve width versus input amplitude for symmetric and antisymmetric modes in "in-phase" modulated array at $\mu=0.2$. (d) Resonance curve width versus input amplitude for symmetric mode in "out-of-phase" modulated array at $\mu=0.2$. The curve for antisymmetric mode is not shown since it almost coincides with that for symmetric mode.

Figure 4. (a) $U_\mathrm{m}$ versus input amplitude $A$ for symmetric and antisymmetric modes in "in-phase" modulated waveguide array at $\mu=0.2$ and $\Omega=\Omega_\mathrm{r}$. Dashed line indicates $U_\mathrm{m}=1$ level. (b) Amplitude of delocalization $A_\mathrm{del}$ versus longitudinal modulation depth $\mu$ for symmetric and antisymmetric modes in "in-phase" modulated waveguide array.

Figure 5. (a) Diffraction of antisymmetric mode in unmodulated array. Dynamics of propagation of antisymmetric mode in "in-phase" modulated waveguide array at $\mu=0.15$ for (b) $A=0.01$, (c) $1.59$, and (d) $2.00$. Dynamics of propagation of (e) symmetric and (f) antisymmetric modes at $\mu=0.15$, $A=0.99$. In panels (b)-(f) $\Omega=\Omega_\mathrm{r}$.

Figure 6. (a) "In-phase" modulated array. Propagation of antisymmetric mode in (b) "out-of-phase" modulated array and (c) "in-phase" modulated array at $\mu = 0.2$. In all cases $A = 0.01$.

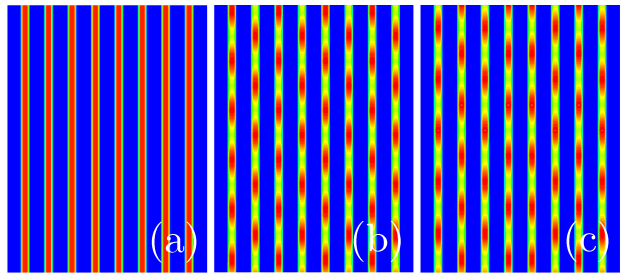

Figure 1.     Unmodulated (a), "out-of-phase" modulated (b), and "in-phase" modulated (c) waveguide arrays.

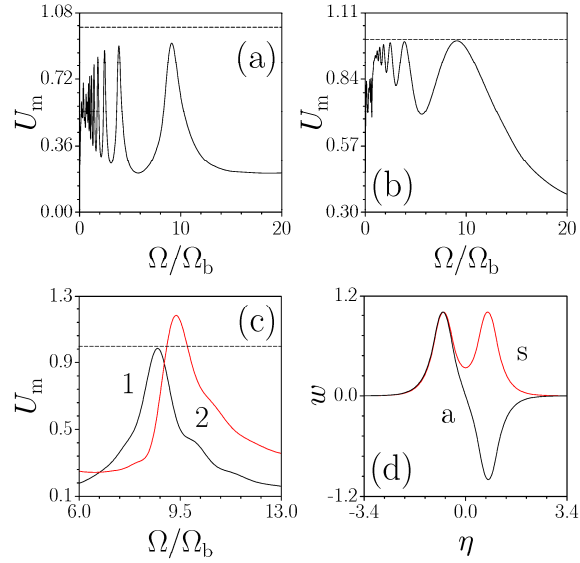

Figure 2. Theoretically calculated $U_\mathrm{m}$ value in both central waveguides versus $\Omega/\Omega_\mathrm{b}$ at $\mu=0.2$ and $A=0.01$ for antisymmetric mode in (a) "out-of-phase" modulated and (b) "in-phase" modulated waveguide arrays. (c) $U_\mathrm{m}$ versus $\Omega/\Omega_\mathrm{b}$ in right (curve 1) and left (curve 2) waveguides for antisymmetric mode in "out-of-phase" modulated array at $\mu=0.2$ and $A=0.01$. Dashed lines in (a)-(c) indicate $U_\mathrm{m}=1$ level. (d) Profiles of symmetric and antisymmetric modes of two waveguides.

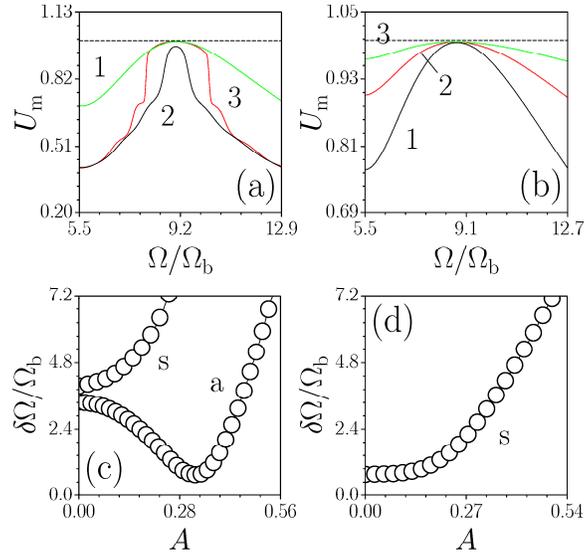

Figure 3. $U_\mathrm{m}$ versus $\Omega/\Omega_\mathrm{b}$ for (a) antisymmetric mode with amplitude $A = 0.01$ (curve 1), 0.35 (curve 2), 0.42 (curve 3) and (b) symmetric mode with amplitude $A = 0.01$ (curve 1), 0.25 (curve 2), 0.40 (curve 3) in "in-phase" modulated waveguide array at $\mu = 0.2$. Dashed lines indicate $U_\mathrm{m} = 1$ level. (c) Resonance curve width versus input amplitude for symmetric and antisymmetric modes in "in-phase" modulated array at $\mu = 0.2$. (d) Resonance curve width versus input amplitude for symmetric mode in "out-of-phase" modulated array at $\mu = 0.2$. The curve for antisymmetric mode is not shown since it almost coincides with that for symmetric mode.

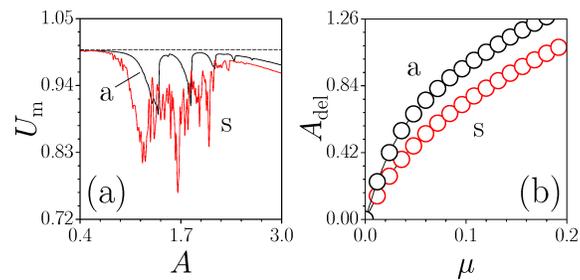

Figure 4.   (a) $U_\mathrm{m}$ versus input amplitude $A$ for symmetric and antisymmetric modes in "in-phase" modulated waveguide array at $\mu = 0.2$ and $\Omega = \Omega_\mathrm{r}$. Dashed line indicates $U_\mathrm{m} = 1$ level. (b) Amplitude of delocalization $A_\mathrm{del}$ versus longitudinal modulation depth $\mu$ for symmetric and antisymmetric modes in "in-phase" modulated waveguide array.

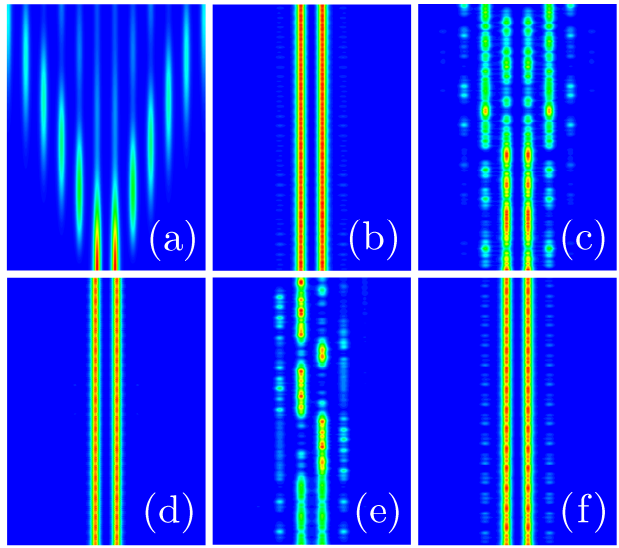

Figure 5. (a) Diffraction of antisymmetric mode in unmodulated array. Dynamics of propagation of antisymmetric mode in "in-phase" modulated waveguide array at $\mu = 0.15$ for (b) $A = 0.01$, (c) $1.59$, and (d) $2.00$. Dynamics of propagation of (e) symmetric and (f) antisymmetric modes at $\mu = 0.15$, $A = 0.99$. In panels (b)-(f) $\Omega = \Omega_\mathrm{r}$.

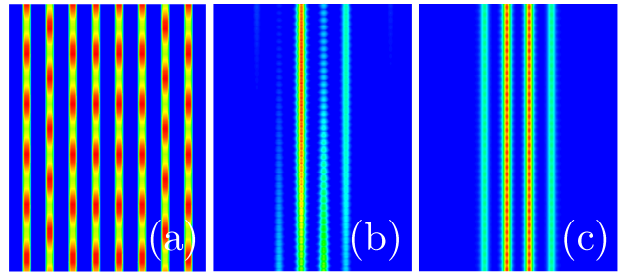

Figure 6. (a) "In-phase" modulated array. Propagation of antisymmetric mode in (b) "out-of-phase" modulated array and (c) "in-phase" modulated array at $\mu = 0.2$. In all cases $A = 0.01$.